\documentclass[12pt,preprint]{aastex}

\newcommand{\kms}{\ensuremath{\rm km\,s^{-1}}}
\newcommand{\ms}{\ensuremath{\rm m\,s^{-1}}}
\newcommand{\gcmc}{\ensuremath{\rm g\,cm^{-3}}}
\newcommand{\teff}{\ensuremath{T_{\rm eff}}}
\newcommand{\logg}{\ensuremath{\log{g}}}
\newcommand{\vsini}{\ensuremath{v \sin{i}}}
\newcommand{\feh}{\rm{[Fe/H]}}

\newcommand{\rsun}{\ensuremath{R_\sun}}
\newcommand{\msun}{\ensuremath{M_\sun}}
\newcommand{\lsun}{\ensuremath{L_\sun}}

\newcommand{\rstar}{\ensuremath{R_\star}}
\newcommand{\mstar}{\ensuremath{M_\star}}
\newcommand{\loggstar}{\ensuremath{\logg_\star}}
\newcommand{\lstar}{\ensuremath{L_\star}}

\newcommand{\rpl}{\ensuremath{R_{\rm P}}}
\newcommand{\mpl}{\ensuremath{M_{\rm P}}}

\newcommand{\rhopl}{\ensuremath{\rho_{\rm P}}}
\newcommand{\loggpl}{\ensuremath{\logg_{\rm P}}}
\newcommand{\teq}{\ensuremath{T_{\rm eq}}}

\newcommand{\rjup}{\ensuremath{R_{\rm J}}}
\newcommand{\mjup}{\ensuremath{M_{\rm J}}}

\newcommand{\koicur}{Kepler-6}
\newcommand{\koicurb}{Kepler-6b}

\newcommand{\koicurCCra}{\ensuremath{19^{\mathrm{h}}47^{\mathrm{m}}20^{\mathrm{s}}.94}}
\newcommand{\koicurCCdec}{\ensuremath{+48^{\circ}14'23''.8}}
\newcommand{\koicurCCkic}{KIC~10874614}
\newcommand{\koicurCCtwomass}{2MASS~19472094+4814238}
\newcommand{\koicurCCkicr}{13.253}			

\newcommand{\koicurLCar}{\ensuremath{7.05^{+0.11}_{-0.06}}}			% a/R*  Jason
\newcommand{\koicurLCrprstar}{\ensuremath{0.09829^{+0.00014}_{-0.00050}}}	% Rp/R* Jason
\newcommand{\koicurLCimp}{\ensuremath{0.398^{+0.020}_{-0.039}}}			% b, Jason
\newcommand{\koicurLCi}{\ensuremath{86\fdg8\pm0.3}}			% inclination, Jason

\newcommand{\koicurLCP}{\ensuremath{3.234723\pm0.000017}}	% Period and error, Jason
\newcommand{\koicurLCPshort}{3.235}				% Period, short 
\newcommand{\koicurLCPprec}{\ensuremath{3.234723}}
\newcommand{\koicurLCT}{\ensuremath{2454954.48636\pm0.00014}}	% Epoch from Jason's model
\newcommand{\koicurSMEteff}{\ensuremath{5647\pm44}}	% HIRES, Debra Fischer
\newcommand{\koicurSMEfeh}{\ensuremath{+0.34\pm0.04}}	% HIRES
\newcommand{\koicurSMEvsin}{\ensuremath{3.0\pm1.0}}	% HIRES

\newcommand{\koicurYYmshort}{\ensuremath{1.21}}				%
\newcommand{\koicurYYmlong}{\ensuremath{1.209^{+0.044}_{-0.038}}}	% Jason's; agrees with Tim
\newcommand{\koicurYYrshort}{\ensuremath{1.39}}				%
\newcommand{\koicurYYrlong}{\ensuremath{1.391^{+0.017}_{-0.034}}}	% 
\newcommand{\koicurYYlogg}{\ensuremath{4.236\pm0.011}}			% log(g*) Jason's value
\newcommand{\koicurYYlum}{\ensuremath{1.99^{+0.24}_{-0.21}}}		% L/Lsun.  Tim's value
\newcommand{\koicurYYage}{\ensuremath{3.8\pm1.0}}			% Age, Gyr.  Tim's value
\newcommand{\koicurRVK}{\ensuremath{80.9\pm2.6}}			% m/s
\newcommand{\koicurRVgamma}{\ensuremath{-18.3\pm3.5}}			% relative in m/s
\newcommand{\koicurRVmean}{\ensuremath{-49.14\pm0.10}}			% Absolute in km/s, not relative
\newcommand{\koicurPPlogg}{\ensuremath{2.974^{+0.016}_{-0.022}}}	% log(g) for planet (cgs)
\newcommand{\koicurPParel}{\ensuremath{0.04567^{+0.00055}_{-0.00046}}}		% a in AU
\newcommand{\koicurPPrho}{\ensuremath{0.352^{+0.018}_{-0.022}}}	% Planet density (gm/cc)
\newcommand{\koicurPPrhoshort}{\ensuremath{0.35}}

\newcommand{\koicurPPm}{\ensuremath{0.669^{+0.025}_{-0.030}}}		%  Jason's
\newcommand{\koicurPPmshort}{\ensuremath{0.67}}				%
\newcommand{\koicurPPr}{\ensuremath{1.323^{+0.026}_{-0.029}}}		%	 Jason's
\newcommand{\koicurPPrshort}{\ensuremath{1.32}}				%
\newcommand{\koicurPPteq}{\ensuremath{1500\pm200}}			% Need Jason's value

\shortauthors{Dunham et al.}
\shorttitle{\koicurb}
\slugcomment{Version 5.2 --- 28 December 2009}

\begin{document}

\title{\koicurb: A Transiting Hot Jupiter Orbiting a Metal-Rich Star\altaffilmark{\dagger}}

\altaffiltext{$\dagger$}
{Based in part on observations obtained at the W.~M.~Keck Observatory,
which is operated by the University of California and the California
Institute of Technology.}

\author{
Edward~W.~Dunham\altaffilmark{1},
William~J.~Borucki\altaffilmark{2},
David~G.~Koch\altaffilmark{2},
Natalie~M.~Batalha\altaffilmark{3},
Lars~A.~Buchhave\altaffilmark{4,5},
Timothy~M.~Brown\altaffilmark{6},
Douglas~A.~Caldwell\altaffilmark{7},
William~D.~Cochran\altaffilmark{8},
Michael~Endl\altaffilmark{8},
Debra~Fischer\altaffilmark{17},
Gabor~F\H{u}r\'{e}sz\altaffilmark{4},
Thomas~N.~Gautier III\altaffilmark{9},
John~C.~Geary\altaffilmark{4},
Ronald~L.~Gilliland\altaffilmark{10},
Alan~Gould\altaffilmark{16},
Steve~B.~Howell\altaffilmark{11},
Jon~M.~Jenkins\altaffilmark{7},
Hans~Kjeldsen\altaffilmark{15},
David~W.~Latham\altaffilmark{4},
Jack~J.~Lissauer\altaffilmark{2},
Geoffrey~W.~Marcy\altaffilmark{12},
Soren~Meibom\altaffilmark{4},
David~G.~Monet\altaffilmark{13},
Jason~F.~Rowe\altaffilmark{14,7},
Dimitar~D.~Sasselov\altaffilmark{4}
}

\altaffiltext{1}{Lowell Observatory, Flagstaff, AZ 86001}

\altaffiltext{2}{NASA Ames Research Center, Moffett Field, CA 94035}

\altaffiltext{3}{San Jose State University, San Jose, CA 95192}

\altaffiltext{4}{Harvard-Smithsonian Center for Astrophysics, Cambridge, MA 02138}

\altaffiltext{5}{Niels Bohr Institute, Copenhagen University, DK-2100 Copenhagen, Denmark}

\altaffiltext{6}{Las Cumbres Observatory Global Telescope, Goleta, CA 93117}

\altaffiltext{7}{SETI Institute, Mountain View, CA 94043}

\altaffiltext{8}{University of Texas, Austin, TX 78712}

\altaffiltext{9}{Jet Propulsion Laboratory/California Institute of Technology, Pasadena, CA 91109}

\altaffiltext{10}{Space Telescope Science Institute, Baltimore, MD 21218}

\altaffiltext{11}{National Optical Astronomy Observatory, Tucson, AZ 85719}

\altaffiltext{12}{University of California, Berkeley, Berkeley, CA 94720}

\altaffiltext{13}{US Naval Observatory, Flagstaff Station, Flagstaff, AZ 86001}

\altaffiltext{14}{NASA Postdoctoral Program Fellow}

\altaffiltext{15}{University of Aarhus, Aarhus, Denmark}

\altaffiltext{16}{Lawrence Hall of Science, Berkeley, CA 94720}

\altaffiltext{17}{Yale University, New Haven, CT 06510}

\begin{abstract}
We announce the discovery of \koicurb, a transiting hot Jupiter
orbiting a star with unusually high metallicity, $\feh =
\koicurSMEfeh$.  The planet's mass
is about 2/3 that of Jupiter, $\mpl = \koicurPPmshort\,\mjup$, and the
radius is thirty percent larger than that of Jupiter, $\rpl =
\koicurPPrshort\,\rjup$, resulting in a density of $\rhopl =
\koicurPPrhoshort\,\gcmc$, a fairly typical value for such a planet.
The orbital period is $P = \koicurLCPshort$ days.  The host star is
both more massive than the Sun, $\mstar = \koicurYYmshort\,\msun$, and
larger than the Sun, $\rstar = \koicurYYrshort$ \rsun.
\end{abstract}


%=====================================================================

\section{INTRODUCTION}
The {\it Kepler} mission was launched on March 6, 2009 to undertake a search
for Earth-size planets orbiting in the habitable zones of
stars similar to the Sun.  {\it Kepler} uses the transit photometry approach
for this task \citep{Borucki:10a}.  {\it Kepler's} commissioning process went
very well and the system is providing data of exceptional photometric
quality \citep{Koch:10a}.  Indeed, the final commissioning data, 9.7
days of science-like observations of 50000 stars selected from the
{\it Kepler Input Catalog} (KIC) \citep{Koch:10a}, were of such high quality
that they are now commonly, if incorrectly, referred to as the
``quarter 0'' data.  A number of {\it Kepler} Objects of Interest (KOIs)
simply fell out of this dataset, including the present object, known
as KOI-17.  Its transit signal is huge by {\it Kepler} standards, and
excellent light curve results were obtained quickly.  The follow-up
spectroscopic observations took somewhat longer since observations
were not planned to be undertaken so soon after commissioning.  As a
result the first data intended for science, 33.5 days of observations of
150000 KIC stars, became available in the meantime and were folded in
with the original test data for the analysis presented here.  The 1.3
day gap between these two sets of data contained one transit of
\koicurb.  \citet{Haas:10} provide additional details regarding the {\it Kepler} 
data used in this analysis.

Because the pace of early {\it Kepler} extrasolar planet discoveries is limited by
the radial-velocity follow-up observations, the first objects to be
announced tend to be those most easily confirmed spectroscopically.
\koicurb\ (= \koicurCCkic, $\alpha =
\koicurCCra, \delta = \koicurCCdec$, J2000, KIC $r = 
\koicurCCkicr$\,mag) is one of these objects.

\section{KEPLER PHOTOMETRY}
The {\it Kepler} science data for the primary transit search mission
are the long cadence data \citep{Jenkins:10a}.  These consist
of sums very nearly 30 minutes in length (with rigorously the same
integration time) of each pixel in the aperture containing the target
star in question.  These data proceed through an analysis pipeline to
produce corrected pixel data, then simple unweighted aperture
photometry sums are formed to produce a photometric time series for
each object \citep{Jenkins:10b}.  The many thousands of photometric
time series are then processed by the transiting planet search (TPS)
pipeline element \citep{Jenkins:10b}.  The candidate transit events
identified by TPS were then vetted by visual inspection.  

The light curves produced by the photometry pipeline tend to show
drifts due to an extremely small, slow focus change \citep{Jenkins:10a}, and
there are also sometimes low frequency variations in the stellar
signal that can make analysis of the transit somewhat problematic.
These low frequency effects can be removed by modest filters that
have only an insignificant effect on the transit signal
\citep{Koch:10b}.  The unfolded and folded light curves for \koicurb ~produced in
this manner are shown in Figure 1.  \footnote{The time series
photometry and radial velocity data, including bisector spans, 
may be retrieved from the MAST/HLSP data archive at 
http://archive.stsci.edu/prepds/kepler\_hlsp.}

Figure 1 illustrates two of the ways we reject false positives based
on the {\it Kepler} light curves alone.  The lower part of the figure shows
the planetary transit (referred to the left ordinate).  The points
marked with $+$ and $\times$ symbols refer to alternating even and odd
transit events.  This aids in detecting eclipsing binary stars with
nearly, but not quite equal primary and secondary minima.  No such deviation is
observed for \koicur.  The upper curve shows the time of secondary
minimum, assuming zero eccentricity.  The vertical scale, on the
right, is magnified and measured in parts per million to help search
for a small secondary minimum that might be due to occultation of an M dwarf
companion.  Again, no such signature is visible.

\section{FOLLOW-UP OBSERVATIONS}
The crucial importance of ground-based follow-up observations was
recognized long ago \citep{Latham:03} and a well-established plan was in place by the
time of launch.  This has evolved somewhat and our current approach is
described by \citet{Gautier:10}.  The key features are: 1) carry out
high resolution reconnaissance spectroscopy to determine the stellar
properties and the star's suitability for radial-velocity work, and to
search for signs of stellar companions; 2) search for faint blended
eclipsing binaries through the centroid shift approach
\citep{Batalha:10} and by means of high resolution imaging; and 3)
detect the stellar reflex motion with precise radial-velocity
measurements.  In the future we hope to improve our
constraints on stellar properties by including both
asteroseismology results for at least the brighter host stars
\citep{Gilliland:10a} and parallaxes \citep{Monet:10} for every target.
These results are not available for \koicur\ at this time.

\subsection{Reconnaissance Spectroscopy}
The reconnaissance spectra for \koicur\ were carried out very early in
the follow-up program, so observations were obtained with several of
our resources for comparison purposes: the TRES spectrograph on the
Tillinghast telescope (by L.~Buchhave and D.~Latham), the McDonald 
2.7-m Coude spectrograph (by M.~Endl and W.~Cochran), and the FIES 
spectrograph on the Nordic Optical Telescope (by L.~Buchhave and G.~F\H{u}r\'{e}sz).  
The spectra indicated that \koicur\ was well suited for precise radial-velocity 
work and no evidence of double lines or a stellar companion was seen.  
The best data on stellar properties were determined later and are discussed
below.

The mean radial velocity of \koicur\ was determined by C. Chubak with 
two HIRES spectra using telluric lines to set their zero point.  Observations 
of the radial velocity standard HD182488 using the same approach 
produced a mean value of $-21.50\pm0.18$\,\kms, in good agreement with 
the adopted value of -21.508\,\kms.  Observations with FIES and TRES 
using the same standard star to determine their zero points were in good 
agreement with the HIRES velocity.  The average of these values is 
shown in Table~\ref{tab:parameters}.

\subsection{High-Resolution Imaging and Centroid Motion}
\koicur\ has a companion $4.1\arcsec$ distant and 3.8 magnitudes
fainter as seen in both a HIRES guide camera image and a PHARO $J$
band AO image obtained with the Palomar Hale telescope (Gautier and Ciardi).  
A WIYN speckle image (Howell) shows only a single star within
its $2\arcsec$ square box.  There are apparently two additional 
components both approximately 5.4 magnitudes fainter and $11.5\arcsec$ 
distant that appear in the KIC as KIC IDs 10874615 and 10874616.  
These two objects refer to a single star and erroneously
appear as two objects in the KIC.  This star is
very close to the edge of the aperture containing \koicur.  
The observed shift in the {\it Kepler} intensity weighted mean image
centroid during transit, ($-0.1$, $+0.3$) millipixels in the (column, row) 
directions is consistent with the expected shift if the faint companions
are not the source of the transit signature.  If the fainter, more distant star
is fully in the aperture the expected shift is ($0.0$, $+0.3$) millipixels and if it
is completely out of the aperture the expected shift is ($-0.1$, $+0.4$) millipixels,
both being consistent with the observed centroid shift.  On the other hand 
if the closer, brighter background star is the transit source the expected shift 
is 30 times larger, ruling out this possibility.  The fainter object is too faint to 
produce the observed transit depth even if it disappears entirely during the
transit, and if it could the resulting centroid shift would be nearly a factor of 100 larger
than the observed shift.  The typical sensitivity level for detection of
centroid shifts is on the order of 0.1 millipixel or somewhat less.
See \citet{Batalha:10} for a detailed discussion of this approach for
candidate vetting.

\subsection{PRECISE RADIAL VELOCITIES}
Precise velocities were obtained by using HIRES on the Keck 1 telescope
with iodine cell radial-velocity reference.  A modified version of the 
standard HIRES iodine cell reduction pipeline that includes improved cosmic ray 
rejection and sky brightness suppression was applied to these spectra by 
J.~Johnson.  The reduction pipeline modification was necessary because
of the faintness of this star compared to the bright stars normally observed
for radial velocity planet search work.  The nearby companion mentioned earlier
is sufficiently far away on the sky that it does not interfere with
the HIRES work.  The phased velocities are shown in
Figure \ref{fig:orbit} with a fit to a circular
orbit whose phase and period are fixed by the observed transit times.

The Monte Carlo analysis described below shows no evidence of orbital
eccentricity.  In this analysis the eccentricity and longitude of periastron 
were parameterized as 
$e \sin \omega$ and $e \cos \omega$.  The probability distribution for these
parameters is symmetrical with zero mean and a standard deviation of 0.029.

The rms of the velocity residuals is $5.7\,\ms$ with the point nearest
transit omitted from the fit.  This point is potentially affected by 
the Rossiter-McLaughlin effect.  Its residual of $12 \,\ms$ is larger than
all the other velocity points and is in the 
sense of a prograde orbit.  The rotational 
velocity given in Table~\ref{tab:parameters} would result in an R-M 
amplitude of $20\,\ms$, certainly detectable with further careful spectroscopy.  
An analysis of the line bisectors 
(Figure \ref{fig:orbit}) shows an rms of $6.8\,\ms$ with no organized structure
that might result from a triple system \citep{Mandushev:05}. 

\section{DISCUSSION}
A Monte Carlo bootstrap procedure developed by Rowe and Brown,
outlined in \citet{Koch:10b} and \citet{Borucki:10b}, that simultaneously fits
the light curve and the radial velocities, and that is consistent with stellar evolution
models, was carried out to produce most of
the stellar and planetary properties presented in
Table~{\ref{tab:parameters}.  The model light curve \citep{Mandel:02} was
integrated over the long cadence integration period prior to fitting to the
{\it Kepler} data.  We note that the nearby companion star
alluded to earlier is only about one {\it Kepler} pixel away from \koicur,
so its light is included in the {\it Kepler} aperture.  The fainter, more distant
star is partially included in the aperture as well.  This has the effect
of diluting the depth of the transit by $3.3\pm0.4$\%.  This effect has been 
taken into account in the modeling process including the dilution uncertainty 
due to the partial inclusion of the more distant star.  The derived planetary density, 
\koicurPPrho\  \gcmc, places it in a well-populated area of the mass-radius relationship for
currently known extrasolar planets.

Several spectra obtained with HIRES without the iodine cell were
subjected to an SME analysis \citep{Valenti:96} by D.~Fischer to provide the remaining
stellar properties reported in Table~\ref{tab:parameters}.  In this analysis
$\logg$ was frozen at the transit-derived value since it is
more reliable than the SME value, which was 4.59 if allowed to
float.  The discrepancy between the transit value of $\logg$ and the SME
value is significant.  \citet{Gilliland:10b} and \citet{Nutzman:10} 
find a similar effect in their
comparison of \logg\ derived from transit fits, asteroseismology
constraints, and SME spectral analysis in the case of HD17156 ($\feh =
+0.24$).  In that case the discrepancy in \logg\ is smaller, 0.12, but
is in the same sense.  \citet{Bakos:09} find a similar situation for HAT-P-13
($\feh = +0.41$) with a \logg\ discrepancy of +0.14.  This suggests that 
there is a systematic error in the \logg\ values derived from SME analysis 
for high metallicity stars.  However, \citet{Burke:07} find only a small \logg\ 
discrepancy of +0.02 in the case of XO-2 ($\feh = +0.45$) and the discrepancy
for HD80606 ($\feh = +0.4$) is only +.01 \citep{Naef:01, Winn:09}.  Further
comparisons of spectroscopic \logg\ values with precise values derived
from transits and asteroseismology are likely to be fruitful.

\acknowledgements 
Funding for this Discovery mission is provided by NASA's Science
Mission Directorate.  

It is with pleasure that we acknowledge the outstanding work done by
the entire {\it Kepler} team during development, launch, commissioning, and
the ongoing operations of the {\it Kepler} mission.  It is an honor and a privilege to
work with such a talented and dedicated group of people.

We thank the anonymous referee for several excellent suggestions that 
improved the paper.

{\it Facilities:} \facility {The Kepler Mission, Tillinghast (TRES), NOT (FIES), McDonald 2.7-m (Coude Spectrograph), Keck:I (HIRES), WIYN (Speckle), Palomar Hale (PHARO/NGS)}

%======================================================================

%====================================================================

% Sample of the photometric data for electronic publication only.
% This can be omitted from the printed version if space is a problem.

%\begin{deluxetable}{cc}
%\tablewidth{0pc}
%\tablecaption{{\it Kepler} Photometry for \koicur\label{tab:phot}}
%\tablehead{
%\colhead{Heliocentric Julian Day} &
%\colhead{Normalized Intensity}
%} 
%\startdata
%2454953.538374 & 1.000068 \\
%2454953.558809 & 0.999986 \\
%2454953.579243 & 0.999977 \\
%2454953.599678 & 0.999826 \\
%2454953.620112 & 0.999816 \\
%\enddata
%\tablecomments{
%The photometric data are fully presented in the electronic edition
%of the Astrophysical Journal. The portion shown here for guidance
%regarding its form and content.} \\
%\end{deluxetable}

%==============================================================================

\begin{deluxetable}{rrrr}
\tablewidth{0pc}
\tablecaption{Relative Radial-Velocity Measurements of \koicur\label{tab:rvs}}
\tablehead{
\colhead{BJD}				&
\colhead{Phase}				&
\colhead{RV}				&
\colhead{\ensuremath{\sigma_{\rm RV}}}	\\ 
% \colhead{(days)}			&
% \colhead{(cycles)}			&
&
&
\colhead{(\ms)}				&
\colhead{(\ms)}
}
\startdata
2454986.087289 & 0.769 &$ +68.42 $& 4.48 \\
2454987.087348 & 0.078 &$ -59.94 $& 4.59 \\
2454988.077695 & 0.384 &$ -67.64 $& 4.75 \\
2454989.019649 & 0.676 &$ +48.64 $& 4.64 \\
2454995.112273 & 0.559 &$ +18.56 $& 4.79 \\
2455014.881155 & 0.671 &$ +51.75 $& 4.54 \\
2455015.984052 & 0.012 &$ -34.79 $& 4.65 \\
2455017.067098 & 0.346 &$ -75.84 $& 4.94 \\
2455044.019956 & 0.679 &$ +50.59 $& 5.42 \\
\enddata
\end{deluxetable}

%========================================================================

\begin{deluxetable}{lcc}
\tabletypesize{\scriptsize}
\tablewidth{0pc}
\tablecaption{System Parameters for \koicur \label{tab:parameters}}
\tablehead{\colhead{Parameter}	& 
\colhead{Value} 		& 
\colhead{Notes}}
\startdata
\sidehead{\em Transit and orbital parameters}
Orbital period $P$ (d)                          & \koicurLCP            & A     \\
Midtransit time $E$ (HJD)                       & \koicurLCT            & A     \\
Scaled semimajor axis $a/\rstar$                & \koicurLCar           & A     \\
Scaled planet radius \rpl/\rstar                & \koicurLCrprstar      & A     \\
Impact parameter $b \equiv a \cos{i}/\rstar$    & \koicurLCimp          & A     \\
Orbital inclination $i$                   & \koicurLCi            & A     \\
Orbital semi-amplitude $K$ (\ms)                & \koicurRVK            & A,B   \\
Orbital eccentricity $e$                        & 0 (adopted)         & A,B   \\
%Orbital eccentricity $e$                        & \koicurEcc           & A,B   \\
Center-of-mass velocity $\gamma$ (\ms)          & \koicurRVgamma        & A,B   \\
\sidehead{\em Observed stellar parameters}
Effective temperature \teff\ (K)                & \koicurSMEteff        & C     \\
Spectroscopic gravity \logg\ (cgs)              & Fixed at transit value        & C     \\
Metallicity \feh                                & \koicurSMEfeh         & C     \\
Projected rotation \vsini\ (\kms)               & \koicurSMEvsin        & C     \\
Mean radial velocity (\kms)                     & \koicurRVmean         & B     \\
\sidehead{\em Derived stellar parameters}
Mass \mstar (\msun)                             & \koicurYYmlong        & C,D   \\
Radius \rstar (\rsun)                           & \koicurYYrlong        & C,D   \\
Surface gravity \loggstar\ (cgs)                & \koicurYYlogg         & C,D   \\
Luminosity \lstar\ (\lsun)                      & \koicurYYlum          & C,D   \\
%Absolute V magnitude $M_V$ (mag)               & \koicurYYmv           & D     \\
Age (Gyr)                                       & \koicurYYage          & C,D   \\
%Distance (pc)                                  & \koicurXdist          & D     \\
\sidehead{\em Planetary parameters}
Mass \mpl\ (\mjup)                              & \koicurPPm            & A,B,C,D       \\
Radius \rpl\ (\rjup, equatorial)                            & \koicurPPr            & A,B,C,D       \\
Density \rhopl\ (\gcmc)                         & \koicurPPrho          & A,B,C,D       \\
Surface gravity \loggpl\ (cgs)                  & \koicurPPlogg         & A,B,C       \\
Orbital semimajor axis $a$ (AU)                 & \koicurPParel         & E     \\
Equilibrium temperature \teq\ (K)               & \koicurPPteq          & F
\enddata
\tablecomments{\\
A: Based on the photometry.\\
B: Based on the radial velocities assuming an elliptical orbit.\\
C: Based on an SME analysis of the HIRES spectra.\\
D: Based on the Yale-Yonsei stellar evolution tracks.\\
E: Based on Newton's version of Kepler's Third Law and total mass.\\
F: Assumes Bond albedo = 0.1 and complete redistribution.  The uncertainty reflects the uncertainties in these assumptions.
}
\end{deluxetable}

%==============================================================================

\begin{figure}
\plotone{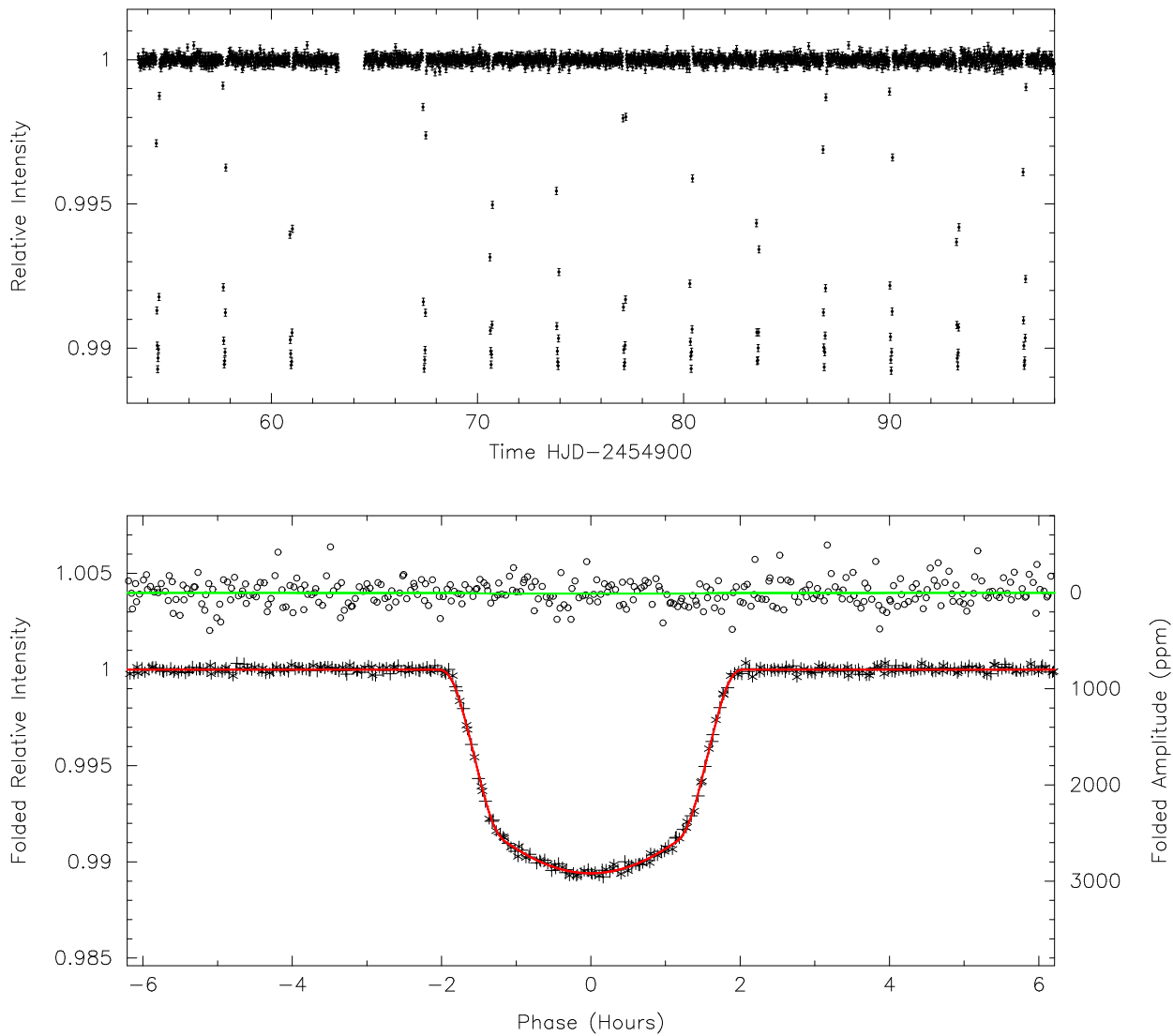}
\caption{
The detrended light curve for \koicur.  The time series for the
entire data set is plotted in the upper panel.  The lower panel shows
the photometry folded by the \koicurLCPprec-day period.  The model fit to
the primary transit is overplotted in red (vertical axis on the left), and our attempt to fit a
corresponding secondary eclipse is shown in green
with the expanded and offset scale on the right.
\label{fig:lightcurve}}
\end{figure}

%==============================================================================

\begin{figure}
\plotone{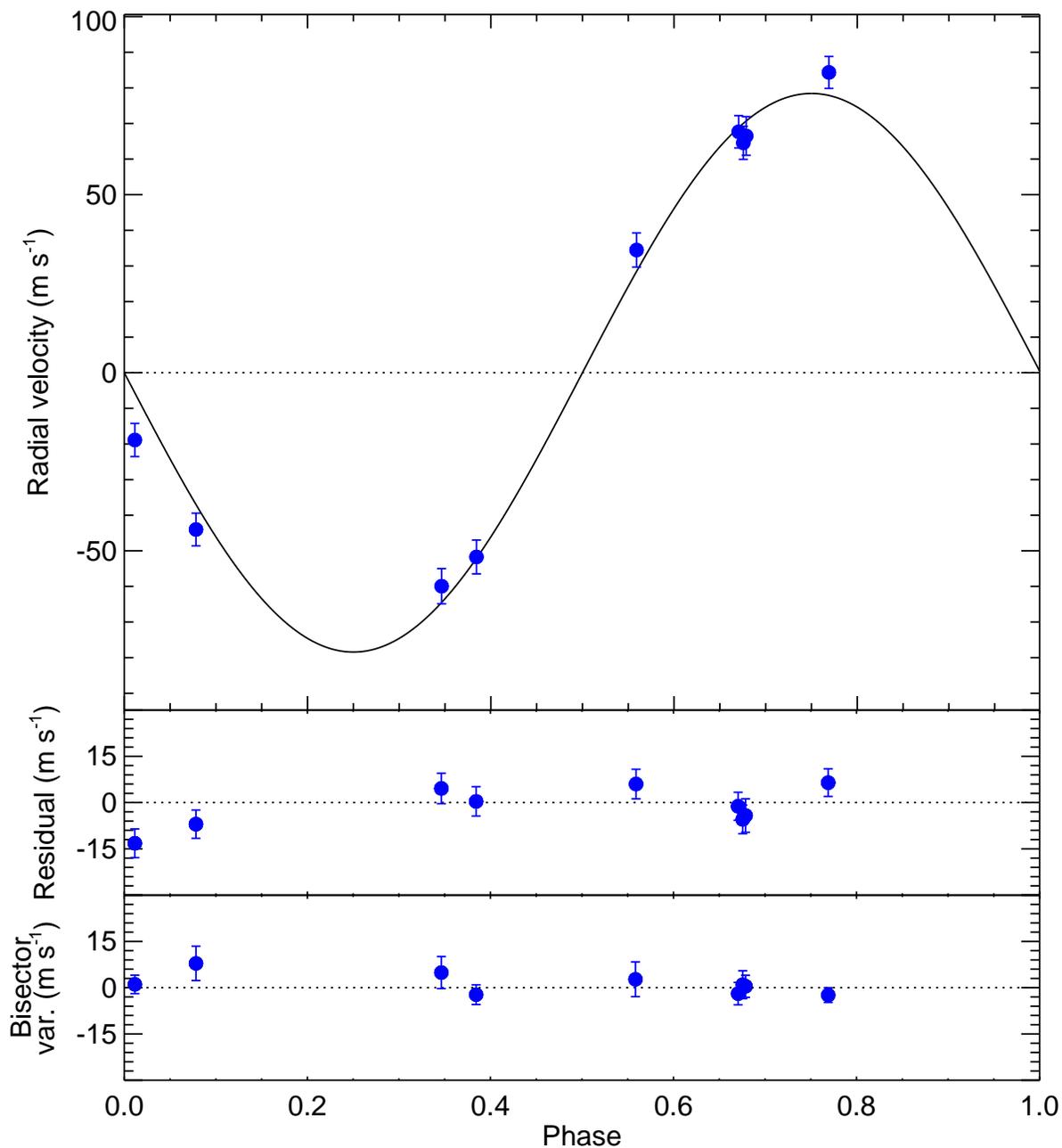}
\caption{
The orbital solution for \koicur. The observed radial velocities
obtained with HIRES on the Keck 1 Telescope are plotted in the top panel
together with the velocity curve for a circular orbit with the period
and time of transit fixed by the photometric ephemeris.  The middle panel shows the 
residuals to the fit.  The point potentially affected by the Rossiter-McLaughlin effect is the
one at the smallest phase.  With this point omitted from the fit the rms of the residuals is 5.7 \ms.
The lower panel shows the line bisector variation.  No structure 
connected to the radial velocity curve is evident.
\label{fig:orbit}}
\end{figure}

\end{document}